\documentstyle[epsfig, twocolumn]{iso98}
\newcommand{\thepapername}{}

\def	\gtsim	{\lower.5ex\hbox{$\; \buildrel > \over \sim \;$}} 
\def	\ltsim	{\lower.5ex\hbox{$\; \buildrel < \over \sim \;$}} 

\def	\beq	{\begin{displaymath}}	
\def	\eeq	{\end{displaymath}}	
\def	\Angstrom	{{\rm\AA}}
\def	\cm	{{\rm \,cm}}
\def	\eV	{{\rm \,eV}}
\def	\kms	{{\rm km\,s}^{-1}}
\def	\K	{{\rm \,K}}
\def	\micron	{{\,\mu{\rm m}}}
\def	\pc	{{\rm\,pc}}
\def	\CII	{{\rm C\,II}}
\def	\FeII	{{\rm Fe\,II}}
\def	\OI	{{\rm O\,I}}
\def	\SiII	{{\rm Si\,II}}
\def	\H	{{\rm H}}
\def	\HH	{{\rm H}_2}
\def	\nH	{n_{\rm H}}

\def	\etal	{{\it et al.}}
\def	\exc	{{\rm exc}}

\def	\gas	{{\rm gas}}
\def	\ISO	{{\it ISO}}

\begin{document}

\setlength{\parindent}{0pt}
\setlength{\parskip}{ 10pt plus 1pt minus 1pt}
\setlength{\hoffset}{-1.5truecm}
\setlength{\textwidth}{ 17.1truecm }
\setlength{\columnsep}{1truecm }
\setlength{\columnseprule}{0pt}
\setlength{\headheight}{12pt}
\setlength{\headsep}{20pt}
\pagestyle{veniceheadings}
\title{
	{\vskip -2.0cm 
	{\normalsize To appear in {\it The Universe As Seen By ISO},
	ed. P. Cox and M. F. Kessler\\
	\ \\
	\ \\
	\ \\
	\ \\}
	}
	{\bf HEATING THE GAS IN PHOTODISSOCIATION FRONTS}
	}
\author{{\bf B. T. Draine$^1$ and Frank Bertoldi$^{2,3}$} \vspace{2mm}\\
$^1$Princeton University Observatory, Princeton, NJ 08544, USA \\
$^2$Max-Planck-Institut f\"ur Radioastronomie, Bonn, Germany\\
$^3$Max-Planck-Institut f\"ur Extraterrestrische Physik, Garching,
	Germany}
\maketitle

\begin{abstract}
	\ISO\ has provided us with a new perspective on gas temperatures
	in photodissociation regions through measurements of line
	emission from rotationally-excited levels of H$_2$.
	The H$_2$ rotational level populations provide a thermal
	probe, showing that gas temperatures $T_{gas}\approx 500-1000\K$
	prevail in a portion of the PDR where significant H$_2$
	is present.
	Such high gas temperatures were unexpected.
	Theoretical models for the S140 PDR are presented.
	Possible mechanisms for heating the gas to such high temperatures
	are discussed.
	\vspace {5pt} \\

	Key~words: ISO; infrared astronomy; molecular hydrogen.
\end{abstract}

\section{INTRODUCTION
	\label{sec:intro}
	}

Photodissociation fronts play an important part in the global energetics
of star-forming galaxies, as an appreciable fraction of the energy
radiated by newly-formed stars is reprocessed in photodissociation fronts,
resulting in $\HH$ and fine structure line emission.
If we want to understand the emission spectra of star-forming galaxies,
good theoretical models of photodissociation fronts are required.

\ISO\ observations of line emission from rotationally-excited levels of
$\HH$ have
provided unequivocal evidence for gas temperatures in the
$500-1000\K$ range in a portion of the photodissociation region (PDR) 
where the $\HH$ fraction is appreciable.
These temperatures were higher than expected from current models of the
heating and cooling processes in PDRs, and therefore require reconsideration
of the physics of the gas and dust in PDRs.

In \S\ref{sec:whatisapdr} we review the basic structure of photodissociation
fronts, and in \S\ref{sec:thermalprobe} we discuss how $\HH$ can be used
as a thermometer to indicate the gas temperatures in PDRs.
In \S\ref{sec:s140} we use the S140 PDR as an example.
Theoretical models for the heating and cooling in PDRs are discussed
in \S\ref{sec:thermal_balance},
and in \S\ref{sec:what_to_change} we stress some of the uncertainties in
the modelling of PDRs.
We summarize in \S\ref{sec:summary}

\section{WHAT IS A PHOTODISSOCIATION FRONT?
	\label{sec:whatisapdr}
	}

The term ``photodissociation front'' refers to the interface separating
a region which is predominantly molecular, and a region where 
the ultraviolet energy density is sufficiently high so that the molecular
fraction is $\ll 1$.
Young O and B stars located near their natal (only partially-disrupted)
molecular clouds produce
conspicuous (i.e., high surface brightness) photodissociation fronts.
A number of reviews of PDRs have appeared recently
(\cite{HoTi97}, \cite{SYD98}, \cite{Wa98}), including a comprehensive
article by \cite*{HoTi98}.

\begin{figure}[ht]
\begin{center}
\leavevmode
\centerline{
	\epsfig{
		file=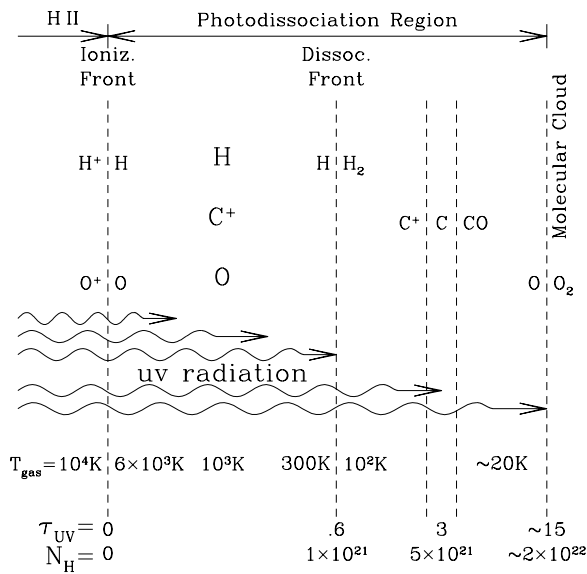,
		width=8.0cm
		}
	}
\end{center}
\caption{\em The structure of a PDR produced by radiation from a star
	which also emits enough ionizing photons to produce an HII region.
	The ionization front separates the HII region from the PDR.
	Typical values of the effective
	far-ultraviolet optical depth, and
	H nucleon column density are shown.
	}
\label{fig:pdrcartoon}
\end{figure}

A photodissociation front is not a mathematical surface, but rather an
extended ``photodissociation region'' (PDR).
Figure \ref{fig:pdrcartoon} is a cartoon illustrating the different
zones within the PDR.

The structure of the PDR is determined primarily by the attenuation of the 
far-ultraviolet (6 - 13.6 eV)
radiation field,
as one moves from the ionization front 
into the PDR.
The $\HH$ abundance results from a balance between formation of $\HH$ on
dust grains and photodissociation of $\HH$ by $912 < \lambda < 1110\Angstrom$
photons.
The photodissociation rate
is determined by both $\HH$ self-shielding (\cite{DrBe96})
as well as attenuation by
dust.
Because of this self-shielding, the H/$\HH$ transition occurs closer to the
ionization front
than the
C$^+$/C/CO
or O/O$_2$ transitions.
Atoms with ionization potentials lower
than H (e.g., C) are photoionized
in the PDR, whereas species with ionization potentials greater than H
(e.g. O) are neutral in the PDR.

Ionization fronts and dissociation fronts 
are, in general, time-dependent structures (\cite{BeDr96}).
Under many circumstances, however, the structure can be approximated
as being stationary in a frame moving with the
dissociation front.
Furthermore, if the dissociation front is advancing into the molecular
gas with a propagation speed 
$v_{\rm DF}\ltsim 0.3~\kms$ (\cite{BeDr96}),
advection terms can be neglected in the equations governing the thermal
and chemical conditions: the temperature is effectively determined by a local
balance between heating and cooling, and abundances by a local balance
between formation and destruction.
For most PDRs this inequality is expected to be satisfied
(\cite{BeDr96}, \cite{StHo98}), although time dependence may be important
for some planetary nebulae (\cite{HoNa95}).

\section{H$_2$ AS A THERMAL PROBE
	\label{sec:thermalprobe}
	}
The $(v,J)$ rovibrational excited states of $\HH$ are populated by
inelastic collisions
\beq
\HH(v_1,J_1)+X \rightarrow \HH(v_2,J_2) + X \nonumber \pm KE ~~~;
\eeq
by UV pumping:
\begin{eqnarray}
\HH(0,J_1)+h\nu &\rightarrow& \HH^*(v_1,J_1\pm 1)\nonumber\\
	&\rightarrow& \HH(v_2,J_1\pm 0,2) + h\nu {\rm (UV)}
	\nonumber\\
	&\rightarrow& \HH(v_3,J_1\pm 0,2,4) + h\nu {\rm (IR)}
	\nonumber\\
	&& ...\nonumber\\
	&\rightarrow& \HH(0,J_1\pm 0,2,4,...)~~~;
	\nonumber\\
	\nonumber
\end{eqnarray}
and by formation on grains:
\beq
{\rm grain} + \H + \H
\rightarrow {\rm grain} + \HH(v,J) + KE ~~~.
\eeq 
At the densities $n_\H\gtsim 10^4\cm^{-3}$ of bright PDRs, collisions
maintain the ($v$=0,~$J$) levels of $\HH$ in approximate thermal equilibrium
for $J\ltsim 5$.
Therefore measurements with \ISO\ of the quadrupole emission line 
intensities from levels $J\ltsim 5$ 
provide a good indicator of the gas temperature, while the emission from the
higher $J$ levels is sensitive to both the gas temperature and density.
We do not expect the level populations to be characterized by a single
``excitation temperature'', but the ``best-fit'' excitation temperature
characterizing a range of $J$ values (e.g., $J$=3--5) should indicate the
approximate
gas temperature in the part of the PDR where these levels
are predominantly excited.

The populations of very high $J$ levels (e.g., $J=15$) are potentially
affected by UV pumping (the quadrupole decay cascade following a UV pump
injects a small amount of $\HH$ into high $J$ states) and
formation on dust grains (some fraction of the newly-formed $\HH$
may be in high $J$ states).
Therefore excitation temperatures characterizing $J\gtsim 10$ may not
reflect gas temperatures.

\section{THE S140 PDR
	\label{sec:s140}
	}

\begin{figure}[h]
\begin{center}
\leavevmode
\centerline{
	\epsfig{
		file=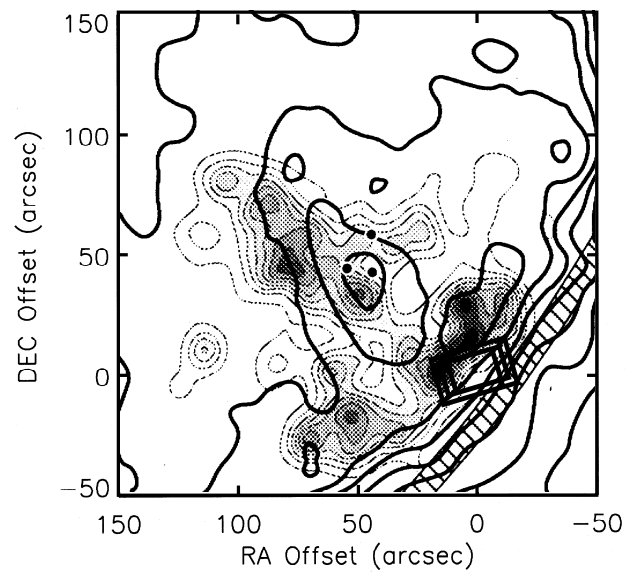,
		width=8.4cm
		}
	}
\end{center}
\caption{\em The S140 PDR, where radiation from the
	B0.5V star HD 211880 impinges on a dense clump in the
	L1202/L1204 molecular cloud.
	The ionization front is shown as a shaded region.
	The greyscale shows [CI]610$\micron$ emission
	(White \& Padman 1991).
	Contours show CO(3-2) emission 
	(Hayashi \etal ~1987).
	The SWS apertures are shown as rectangles positioned just
	interior to the ionization front, where the photodissociation
	front is expected to be located, viewed approximately edge-on.
	The exciting star is located $\sim$7 arcmin SW of the aperture.
	Figure taken from 
	Timmermann \etal (1996).
	}
\label{fig:s140map}
\end{figure}

\ISO\ obtained spectra of
the photodissociation region where the S140 HII region abuts the L1202/L1204
molecular cloud (\cite{TiBe96}); see Figure \ref{fig:s140map}.

\begin{figure}[ht]
\begin{center}
\leavevmode
\centerline{
	\epsfig{file=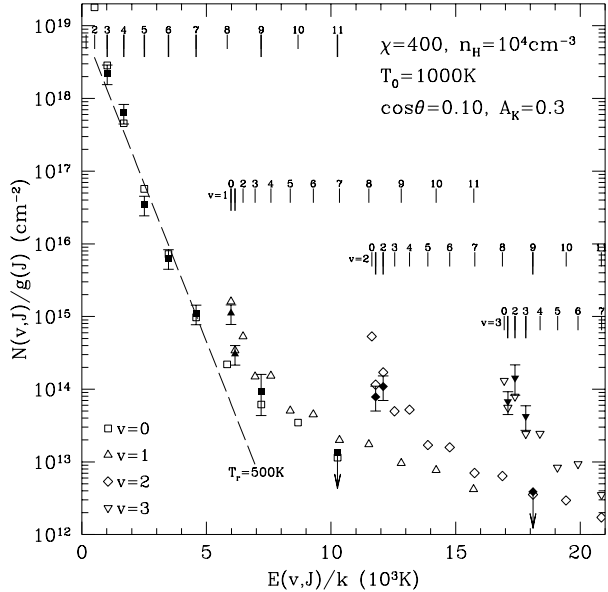,
		width=8.0cm
		}
	}
\end{center}
\caption{\em Observed (filled symbols) and model (open symbols) column
	densities in rovibrational levels of $\HH$ along the line-of-sight.
	Observed intensities have been corrected for foreground extinction by
	dust with 0.3 mag of extinction at $K$.
	The PDR is assumed to be inclined with $\cos\theta=0.1$.
	The broken line is a $T=500\K$ thermal distribution, indicating
	that a significant amount of $\HH$ is located in regions where
	$T_{gas}\gtsim 500\K$.
	Taken from Timmermann \etal\ (1996).}
\label{fig:s140spec}
\end{figure}

\begin{figure}[h]
\begin{center}
\leavevmode
\centerline{
	\epsfig{file=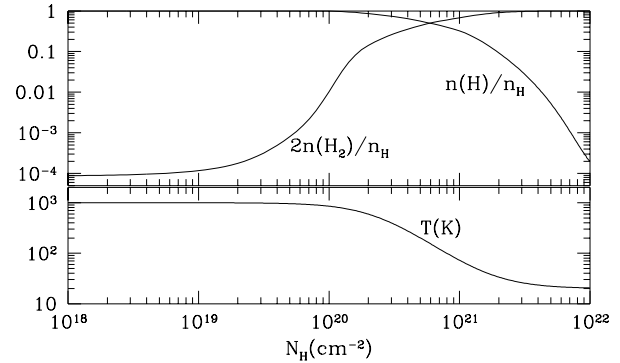,
		width=8.0cm
		}
	}
\end{center}
\caption{\em Ad-hoc temperature profile adopted for S140 PDR model.
	With this assumed temperature profile, the $\HH$
	abundances are computed by requiring a balance between
	$\HH$ formation and photodissociation.
	Taken from Timmermann \etal (1996).}
\label{fig:s140adhocTprof}
\end{figure}

\cite*{TiBe96} reported measurements of line emission out of (0,$J$) levels
with $J$ as large as 9 (see Figure \ref{fig:s140spec}).
The inferred column densities for $3\leq J\leq7$ are characterized by
an excitation temperature 
$T_\exc\approx500\K$; 
the emission from
$J=8$ and 9 suggest a higher temperature.
How much of the PDR is characterized by such high temperatures?
To address this question, \cite*{TiBe96} considered an {\it ad-hoc}
temperature profile, shown in Figure \ref{fig:s140adhocTprof}.

With this {\it ad-hoc} temperature profile,
the chemical abundances are computed by requiring steady-state balance
between formation and destruction, using standard assumptions for the
formation rate of $\HH$ on grains, and a detailed treatment of
$\HH$ self-shielding.
$\HH$ level populations were computed for a plane-parallel PDR,
including both UV pumping and
collisional excitation and deexcitation (\cite{DrBe96}).
The H nucleon density was assumed to be $\nH=10^4\cm^{-3}$
(as expected for pressure balance with the HII region), and
the radiation field at $1000\Angstrom$ was assumed to be
$\chi=400$ times stronger than the \cite*{Ha68} value
(as expected for the $\sim1.7\pc$ distance from the exciting B0.5V star).

The resulting model is shown in Figure \ref{fig:s140spec}, where it is assumed
that $\cos\theta=0.1$, where $\theta$ is the angle between the line-of-sight
and the normal to the model PDR.
It is seen that the level populations are in fairly good agreement with
observations, although the model overestimates the $v=1$ and $2$
level populations, while underestimating the $v=3$ levels.
For this model, $T=500\K$ at the point where $y\equiv 2n(\HH)/\nH=0.2$, and
$T=150\K$ where $y=0.5$.

\cite*{TiBe96} also compared predicted and observed fine structure line
intensities with those predicted by the model.  Agreement is generally
good, allowing for uncertainties in abundances of Si and Fe:
[CI]609$\micron$ is close to the observed value, while 
[SiII]35$\micron$ and [FeII]26$\micron$ agree with observations if
Si/H=$1.5\times10^{-6}$ and Fe/H=$5\times10^{-8}$.
The principal discrepancy is
[OI]63$\micron$, which the model predicts to be $\sim5$ times stronger
than observed.

\section{THERMAL BALANCE IN PDRS
	\label{sec:thermal_balance}
	}

The obvious question now is: what temperatures do we {\it expect}
the gas to have in PDRs, if we assume that the gas temperature is
determined by a balance between heating and cooling?
The principal heating processes in PDRs are:
\begin{itemize}
\item Photoelectrons emitted from dust grains;
\item UV pumping, followed by collisional deexcitation of 
	rovibrationally excited $\HH$;
\item Photodissociation (kinetic energy of H+H);
\item $\HH$ formation (kinetic energy, collisional deexcitation);
\item Photoionization of H, C, etc.
\end{itemize}
while the dominant cooling processes are:
\begin{itemize}
\item fine-structure line emission from $[\CII]158\micron$,\\
$[\OI]63,146\micron$,  $[\SiII]35\micron$,
$[\FeII]25\micron$,...;
\item $\HH$ quadrupole emission;
\item CO rotational emission.
\end{itemize}

The grain photoelectric heating rate is quite uncertain.
Here we show results assuming the photoelectric heating rate of
\cite*{BaTi94}.

The cross sections for collisional excitation and deexcitation of $\HH$
are also uncertain; for $T_\gas>1000\K$ we use the H-$\HH$ inelastic rates of
\cite*{MarM95} and \cite*{ManM95}, while for $T_\gas<1000\K$ we extrapolate the
Mandy \& Martin rates using eq. (16) of \cite*{DrBe96},
but using $\theta=1000\K$ rather than $600\K$ as the transition temperature,
as discussed by \cite*{DrBe99}.

\begin{figure}[h]
\begin{center}
\leavevmode
\centerline{
	\epsfig{file=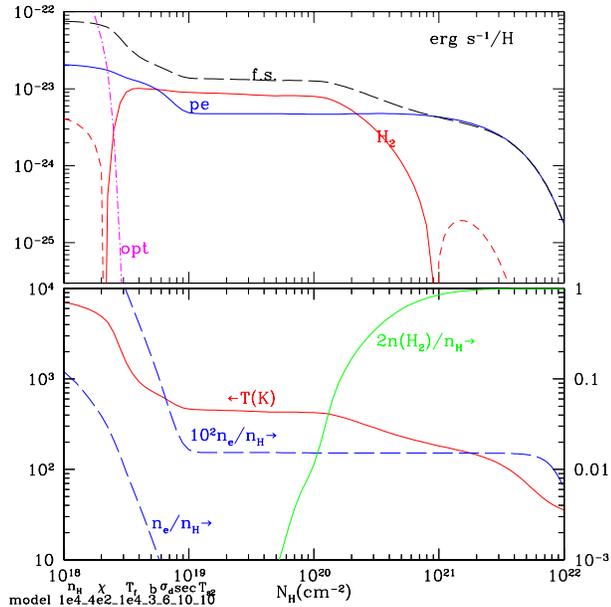,
		width=8.0cm}
		}
\end{center}
\caption{\em Thermochemical profile of S140 PDR model, with temperatures
	determined by thermal balance calculations.
	Lowel panel: gas temperature $T$, $\HH$ fraction
	$2n(\HH)/\nH$, and fractional ionization $n(e)/\nH$.
	Upper panel: heating (solid lines) and cooling (broken lines)
	due to grain photoelectric heating (``pe''), the $\HH$
	molecule  (including formation, collisional excitation and
	deexcitation, and photodissociation),
	fine structure line emission (``f.s.''), and
	optical line emission (``opt'').
	Note that $\HH$ has a net heating effect in much of the PDR.
	}
\label{fig:s140aprof}
\end{figure}

\begin{figure}[h]
\begin{center}
\leavevmode
\centerline{
	\epsfig{
		file=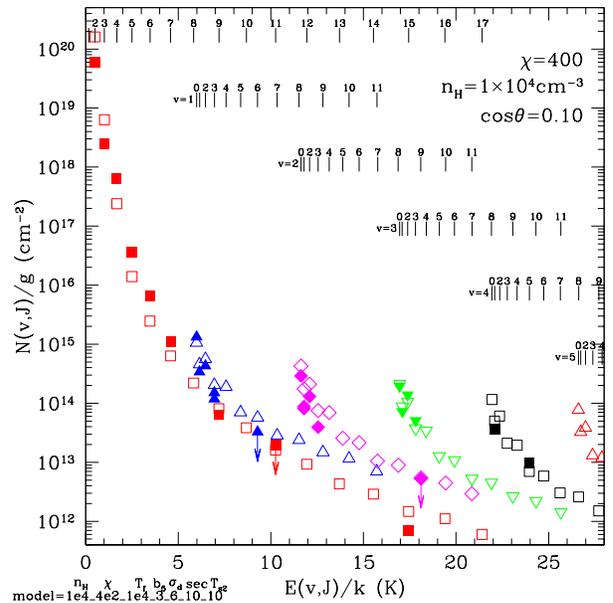,
		width=8.0cm
		}
	}
\end{center}
\caption{\em
	Column densities $N(v,J)$ along the line-of-sight,
	divided by level degeneracies $g(J)$.
	Solid symbols: observed values based on line intensities
	corrected for foreground extinction with
	$A_K=0.2 {\rm mag}$, and internal extinction within the PDR,
	inclined with with $\cos\theta=0.1$.
	Open symbols: model for S140 PDR with $\chi=400$ and
	$\nH=10^4\cm^{-3}$, with temperatures
	determined by thermal balance.
	}
\label{fig:s140aspec}
\end{figure}

The fine-structure line emission is computed using standard collision rates,
abundances characteristic of the gas toward $\zeta$Oph (e.g., Si and Fe depleted
by factors of 40 and 250, respectively).
Some of the lines -- in particular, [OI]$63\micron$ -- can become optically
thick.
An ``escape probability'' approximation was used to estimate
local cooling rates (\cite{DrBe99}).

Figure \ref{fig:s140aprof}
shows the calculated thermochemical
profile, as well as showing the contributions of different processes to
heating and cooling.
Close to the ionization front the fractional ionization is a few percent,
and $T_\gas\approx 4000-6000\K$; in this region, optical line emission
(e.g., [SII]6716,6739) dominates the cooling, but is unimportant
when $T_\gas\ltsim 4000\K$.
Emission from excited fine-structure levels dominates the cooling for
$50\ltsim T_\gas \ltsim 4000\K$.

$\HH$ contributes net heating throughout
much of the PDR.
This heating arises mainly from collisional deexcitation
of rovibrationally excited $\HH$ following UV-pumping or $\HH$ formation on
grains, with a secondary contribution from the kinetic
energy of newly-formed $\HH$ and H+H following photodissociation.
The other major heating process is photoelectric emission from dust grains.
Both $\HH$ heating and photoelectric heating are important for
$N_\H\ltsim 10^{21}\cm^{-2}$; beyond this point $\HH$ self-shielding has
reduced the photodissociation rate (and therefore the $\HH$ formation rate
as well) to very low levels,
and photoelectric heating dominates.

Figure \ref{fig:s140aspec} shows the observed column densities
of excited states of $\HH$ along the line of sight toward the
S140 PDR, corrected for an assumed extinction $A_K=0.3$ mag.
A number of $\HH$ line detections and upper limits were obtained
subsequent to preparation of the \cite*{TiBe96} paper on the S140 PDR;
Figure \ref{fig:s140aspec} shows column densities
obtained from the most recent data reduction (\cite{Beea98}).
Also shown are the column densities predicted by
our model, which assumes an inclination of the PDR with
$\cos\theta=0.1$ -- a factor of 10 ``limb-brightening''.

There are significant differences between the model $\HH$
level populations and those observed toward S140.
The assumed factor of 10 limb-brightening leads to
a good match for
the vibrationally excited levels. However, the
$v=0$ column densities
predicted by the model fall off too 
steeply -- while the $J=2,3$ column densities in the model exceed the
observed values by a factor $\sim 2$, the $J=3-7$ column densities
are too low by factors $\sim 2-3$.
This presumably indicates that either the $\HH$ inelastic cross sections
or the thermal profile are incorrect.
Note, however, that the model appears to be able to reproduce the observed
population of $\HH$ in $J=15$, and thereby a high excitation temperature
for the high$-J$ levels.

\begin{figure}[ht]
\begin{center}
\leavevmode
\centerline{
	\epsfig{file=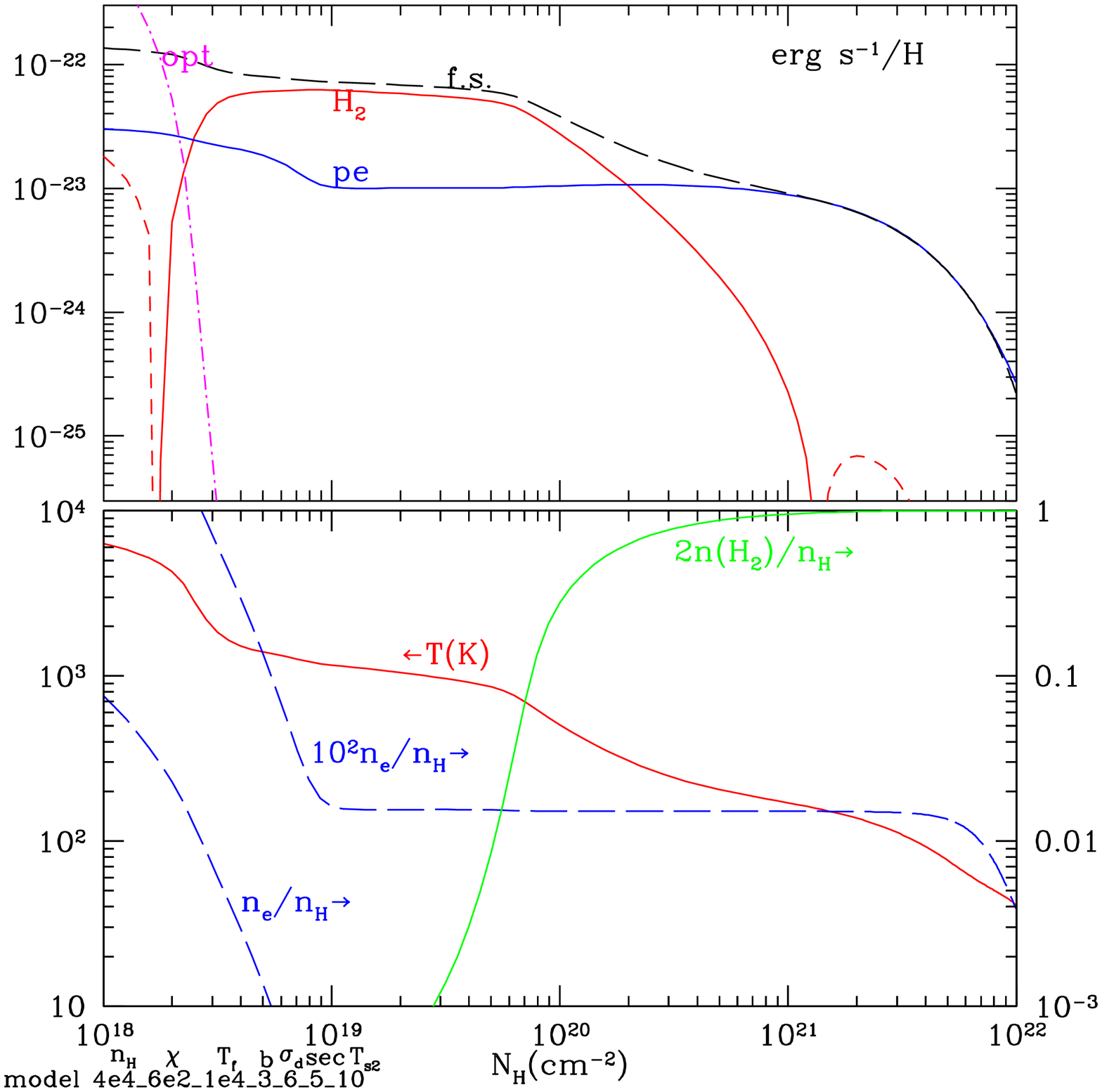,
		width=8.0cm}
	}
\end{center}
\caption{\em Same as Fig.\ \protect{\ref{fig:s140aprof}}, but for
$\nH=4\times10^4\cm^{-3}$
and $\chi=600$.
}
\label{fig:s140bprof}
\end{figure}

\begin{figure}[ht]
\begin{center}
\leavevmode
\centerline{
	\epsfig{file=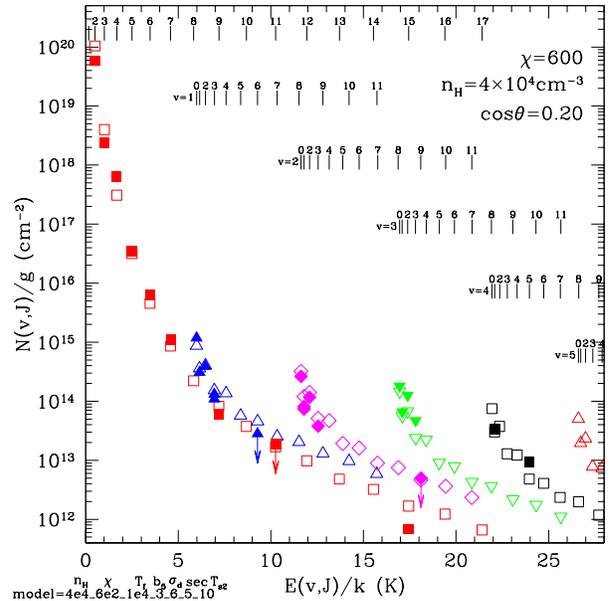,
		width=8.0cm}
	}
\end{center}
\caption{\em Same as Fig.\ \protect{\ref{fig:s140aspec}}, but for
$\nH=4\times10^4\cm^{-3}$ and $\chi=600$, and with
the PDR assumed to be inclined with $\cos\theta=0.2$.
}
\label{fig:s140bspec}
\end{figure}

Somewhat better agreement can be obtained if the gas density and
illuminating radiation field are both increased; in Figures
\ref{fig:s140bprof} and \ref{fig:s140bspec} we show the
result of increasing the
illuminating radiation field to $\chi=600$,
and raising the gas density to $\nH=4\times10^4\cm^{-3}$.
With the increased $\chi$ and $\nH$,
a good fit to the observed $\HH$ line intensities
is obtained for
a limb-brightening factor $1/\cos\theta=5$.
However, the assumed value of $\chi$ is somewhat larger than
expected from our knowledge of the exciting star,
and $\nH$ is larger than expected given our estimates of the
pressure in the ionization front (\cite{TiBe96}).
It should also be noted that even with these values of $\chi$ and $\nH$,
there are still noticeable discrepancies between model and observation.
It is difficult to match the low observed
line ratio between 0-0S(1) and 0-0S(2). Both lines are in the same size
\ISO\ aperture, so beam dilution differences could not be the cause
of the mismatch. The model also overproduces $J=15$, although this
line was observed after \ISO's helium boiloff, which made its
flux calibration more difficult.
Most other levels' column densities are well reproduced.

The high gas temperatures found in S140 are not
unique.
Early evidence for high gas temperatures in PDRs was obtained from 
$\HH$ $J=3$--1 and 4--2 line observations by
\cite*{PaLa91}, who inferred $T_\gas=1000\pm250\K$ at a position .0244~pc
from the ionization front in the Orion Bar, and
$T_\gas=600\pm50\K$ another .0044~pc farther away from the ionization front.

High gas temperatures near the H/$\HH$ transition were 
inferred, based on ground-based spectra, for the reflection nebula NGC 2023
(\cite{DrBe96}).
\ISO\ spectra of the reflection nebula NGC 7023 
(\cite{NGC7023})
and the Orion Bar
(\cite{Roea99}),
also show that some of
the $\HH$ must be in a region of gas temperatures $T_\gas\gtsim600\K$.

\section{WHAT NEEDS TO BE CHANGED IN THE MODELS?
	\label{sec:what_to_change}
	}

Where do we stand on modeling the $\HH$ emission from PDRs?
Considering that the emission arises from a region
with a range of densities,
one might take the point of view that we are doing pretty well -- our
constant density planar PDR
models can approximately reproduce the observed level populations using
ultraviolet pumping of $\HH$ plus collisional excitation/deexcitation,
with gas temperatures obtained from a balance of heating and cooling.
Nevertheless, it seems to us that the discrepancies between our
best models and the observations are unacceptably large.
The discrepancies may be due in part to our underlying 
assumptions of planar geometry
and uniform density, but may also be due to poor approximations in our
description of the local microphysics of heating, cooling, chemistry, and
level excitation. 
Below we indicate some of the areas where existing models may require
modification.

\subsection{$\HH$ inelastic collision cross sections}
The H--$\HH$
collisional excitation/deexcitation rates used here for $T_\gas\gtsim600\K$
are those of \cite*{MarM95} and \cite*{ManM95}, which were obtained from 
quasi-classical trajectory calculations.
These rates are thought to be accurate for $T_\gas\gtsim600\K$,
but are likely to be increasingly inaccurate at low temperatures,
where quantum calculations are required (\cite{Flow97}, \cite{FoBa97}).
\cite*{FoBa97} have recently reported accurate rates
for $\Delta J=\pm2$ transitions among the $v=0$, $J=0-5$ levels,
for $100 < T_\gas < 1000\K$; our extrapolation 
from the \cite*{ManM95} results
was adjusted to give good agreement with these newer rates.

While H-$\HH$ inelastic rates are generally larger than for $\HH$-$\HH$
collisions, the latter are important when the H fraction becomes
small.
Quantum calculations have recently been reported by \cite*{Flow97}.

\subsection{Grain Photoelectric Heating Rate}

The thermal models reported here employ a photoelectric heating rate
recommended by \cite*{BaTi94} for heating by 6-13.6 eV photons:
\beq
\Gamma_{pe}=n_\H\sigma_{uv} u_{uv} c~ \epsilon
\eeq
\begin{eqnarray}
\sigma_{uv} &=& {\rm ~dust ~EUV ~absorption/H }\approx6\times10^{-22}\cm^2
\nonumber\\
u_{uv} &=& {\rm ~EUV ~energy ~density} = \chi\times 
0.033\eV\cm^{-3}\nonumber\\
\epsilon &=& {\rm ~photoelectric ~heating ~efficiency}\nonumber\\
&=& .049(1. + x/1925)^{-0.73} + \nonumber\\
 && .037~ T_4^{0.7}(1. + x/5000)^{-1}\nonumber\\
x&\equiv& \chi (T/\K)^{0.5}/(n_e/\cm^{-3})\nonumber\\
\nonumber
\end{eqnarray}
Bakes \& Tielens assumed a ``MRN'' power-law size distribution down
to ultrasmall grains containing only $\sim30$ carbon atoms
(i.e., PAH molecules).
Since we have only an imprecise understanding of the composition and
size distribution of the grains, this heating rate is necessarily
uncertain.
Under the conditions of interest (e.g., the point where $2n(\HH)/\nH=0.1$ in
Figure \ref{fig:s140aprof}), we have $\chi\approx 400$,
$n_e\approx 1.5\cm^{-3}$, and $T\approx 350\K$, so that $x\approx 5000$,
and the heating efficiency $\epsilon\approx 0.021$, which increases
further into the cloud to $\epsilon\approx 0.049$. Since photoelectric
emission from grains is a dominant heating process in PDRs, 
its further study would be of value.

\subsection{Dust-to-Gas Ratio}

The dust grains play a crucial role in photodissociation regions.
In particular, they determine the rate for formation of $\HH$ from H,
and, as seen above, photoelectric emission from dust is a major
heat source.
Common practice is to adopt a ``standard'' dust-to-gas ratio and
wavelength-dependent dust opacity in the PDR.
However, \cite*{WeDr99} have pointed out that the
anisotropic radiation in a PDR can drive the grains through the gas
at drift velocities which may be comparable to the flow velocity of the
gas relative to the dissociation front, thereby leading to an {\it increase}
in the dust-to-gas ratio in the PDR, at least for those grain compositions
and sizes for which the grain-gas drift velocities are large.
Increased dust concentrations in part of the PDR could alter the
thermochemical properties of the PDR by increasing the $\HH$ formation
rate (and therefore the local rate of UV pumping and
photodissociation of $\HH$) and by increasing the grain photoelectric
heating rate.
As a result, we can expect that the gas temperature will be increased
near the photodissociation front.

The anisotropic radiation incident on a dust grain results in a force
which consists of the ``ordinary'' radiation pressure force due
to absorption and scattering, plus additional forces due to anisotropic
photoelectron emission (the ``illuminated'' side of the grain will
emit more photoelectrons per second than the ``dark'' side) and
anisotropic photodesorption (adsorbed H or $\HH$ can be photodesorbed,
with a larger rate on the ``illuminated'' side of the grain).
The recoil force from photoelectric emission and photodesorption can
be several times larger than the ``ordinary'' radiation pressure
(\cite{WeDr99}).

Because the grains are charged, drift across the magnetic field lines will be
inhibited.
It will be of great interest to see the extent to which these effects
will change the dust-to-gas ratio, and the grain size distribution,
in the PDR, and how this affects the thermochemical profile of the PDR.

\section{SUMMARY
	\label{sec:summary}
	}

\ISO\ has opened a new window on photodissociation fronts, by allowing us
to measure the populations of the rotationally-excited levels of the
$v=0$ state of $\HH$, thereby placing strong constraints on
the gas temperature.
The S140 PDR provides an excellent illustration of this capability.

The observed line emission from rotationally-excited levels of $\HH$
indicates that gas temperatures $T_\gas\gtsim500\K$ prevail in part
of the PDR where appreciable $\HH$ is present.
Attempts to construct
theoretical models of the PDR tend to fall short
in terms of the populations of levels $J=3-7$,

We discuss the uncertainties in the dominant heating and cooling mechanisms;
the $\HH$ inelastic collision
rates still have significant uncertainties,
and the grain photoelectric heating rate is yet not well-established.
In addition, it appears possible that the dust-to-gas ratio in PDRs such
as S140 may deviate from the ``standard'' value since the anisotropic
radiation field can drive the grains through the gas with significant
drift speeds.

\section*{ACKNOWLEDGEMENTS}

For their contribution to the recent S140 data analysis we are grateful to
the MPE SWS group, and especially to H.~Feuchtgruber and E.~Wieprecht.
This research was supported in part by NSF grant AST 96-19429 (BTD)
and by the Deutsche Forschungsgemeinschaft (FB).
B.T.D. wishes to thank Osservatorio Arcetri for its gracious hospitality
during the completion of part of this work.

\section*{}

\end{document}